\documentclass[a4paper]{jpconf}
\usepackage{graphicx,type1cm,eso-pic,color}
\usepackage{color}
\usepackage{url}
\usepackage{xspace}

\begin{document}


\def\shoal{{\it Shoal}\xspace}

\title{Dynamic web cache publishing for IaaS clouds using Shoal}

\author{Ian Gable$^\alpha$, 
Michael Chester$^\alpha$,
Patrick Armstrong$^\beta$,
Frank Berghaus$^\alpha$,
Andre~Charbonneau$^\gamma$,
Colin Leavett-Brown$^\alpha$, 
Michael Paterson$^\alpha$, 
Robert Prior$^\alpha$, 
Randall Sobie$^{\alpha,\delta}$, 
Ryan Taylor$^\alpha$}

\address{$\alpha$: Department of Physics and Astronomy, University of Victoria, Victoria, Canada\\
$\beta$: Computation Institute, University of Chicago, Chicago, USA\\
$\gamma$: Shared Services Canada, Ottawa, Canada\\
$\delta$: Institute of Particle Physics of Canada
}

\ead{igable@uvic.ca}

\begin{abstract}

We have developed a highly scalable application, called \shoal, for tracking and 
utilizing a distributed set of HTTP web caches. 
Our application uses the Squid HTTP cache.
Squid servers advertise their existence to the \shoal server via AMQP messaging by running \shoal
Agent.
The \shoal server provides a simple REST interface that allows clients to
determine their closest Squid cache.
Our goal is to dynamically instantiate Squid caches on IaaS
clouds in response to client demand.
\shoal provides the VMs on IaaS clouds with the location of the nearest 
dynamically instantiated Squid Cache.
In this paper, we describe the design and performance of \shoal. 
 
\end{abstract}

\section{Introduction}

The CERN Virtual Machine File System (CVMFS)~\cite{ref:cvmfs}
is widely adopted by the High Energy Physics (HEP) community for
the distribution of project software.
CVMFS is a read-only network file system that provides access to files from a CVMFS
Server over HTTP.
When CVMFS is used on a cluster of worker nodes, a HTTP web proxy 
can be used to cache the file system contents, so that all subsequent requests 
for that file will be delivered from the local HTTP proxy server. 
Typically, a HEP computing site has a local or regional Squid HTTP 
web proxy~\cite{ref:squidproxy}, with the central CVMFS servers located at 
the main laboratory, such as CERN for the LHC experiments.

The use of IaaS cloud resources is becoming a realistic solution for
HEP workloads~\cite{ref:atlascloud, ref:batchcloud}, and CVMFS is an
effective means of providing the software to the virtual machines (VMs).
Each VM has a list of the available Squid servers and, in most cases,
the Squids are remote.
The optimal Squid may be different depending on the location of the cloud.
Further, one can imagine dynamically instantiating Squid servers in
an opportunistic cloud environment to meet application demand.
However, there is currently no mechanism for locating the optimal Squid server.
As a result, we have developed \shoal as a new service that can dynamically
publish and advertise the available Squid servers.
\shoal is ideal for an environment using both static and dynamic
Squid servers.

\section{Design}

\shoal is divided into three logical modules, a server, an agent, and a client. 
Each package is uploaded to the Python Package Index~\cite{ref:pypi} 
(the standard method of distributing new components in the Python language). 

Each component is designed to provide the functionality of different parts of the system as follows:

\begin{description}

\item[Shoal Server]- is responsible for the following key tasks:
\begin{enumerate}
\item Maintaining a list of active Squid servers in volatile memory and handling AMQP 
messages sent from active Squid servers.
\item Providing a RESTful interface for \shoal Clients to retrieve a list of geographically 
closest Squid servers.
\item Providing basic support for Web Proxy Auto-Discovery Protocol (WPAD).
\item Providing a web user interface to easily view Squid servers being tracked.
\end{enumerate}

\item[Shoal Agent]- is a daemon process run on Squid servers to send an Advanced 
Message Query Protocol (AMQP)~\cite{ref:amqp} message to \shoal Server on a set interval. 
Every Squid server wishing to publish its existence runs \shoal Agent on boot. 
\shoal Agent sends periodic heartbeat messages to the \shoal Server (typically every 30 seconds).

\item[Shoal Client]- is used by worker nodes to query \shoal Server to retrieve
a list  of geographically nearest Squid servers via the 
REST\footnote{Representational State Transfer (REST) is a style of software architecture 
for distributed systems such as the World Wide Web.} interface.
\shoal Client is designed to be simple (less than 100 lines of Python) with no dependencies 
beyond a standard Python installation.

\end{description}

\begin{figure}[t]
\begin{center}
\includegraphics[width=0.6\textwidth]{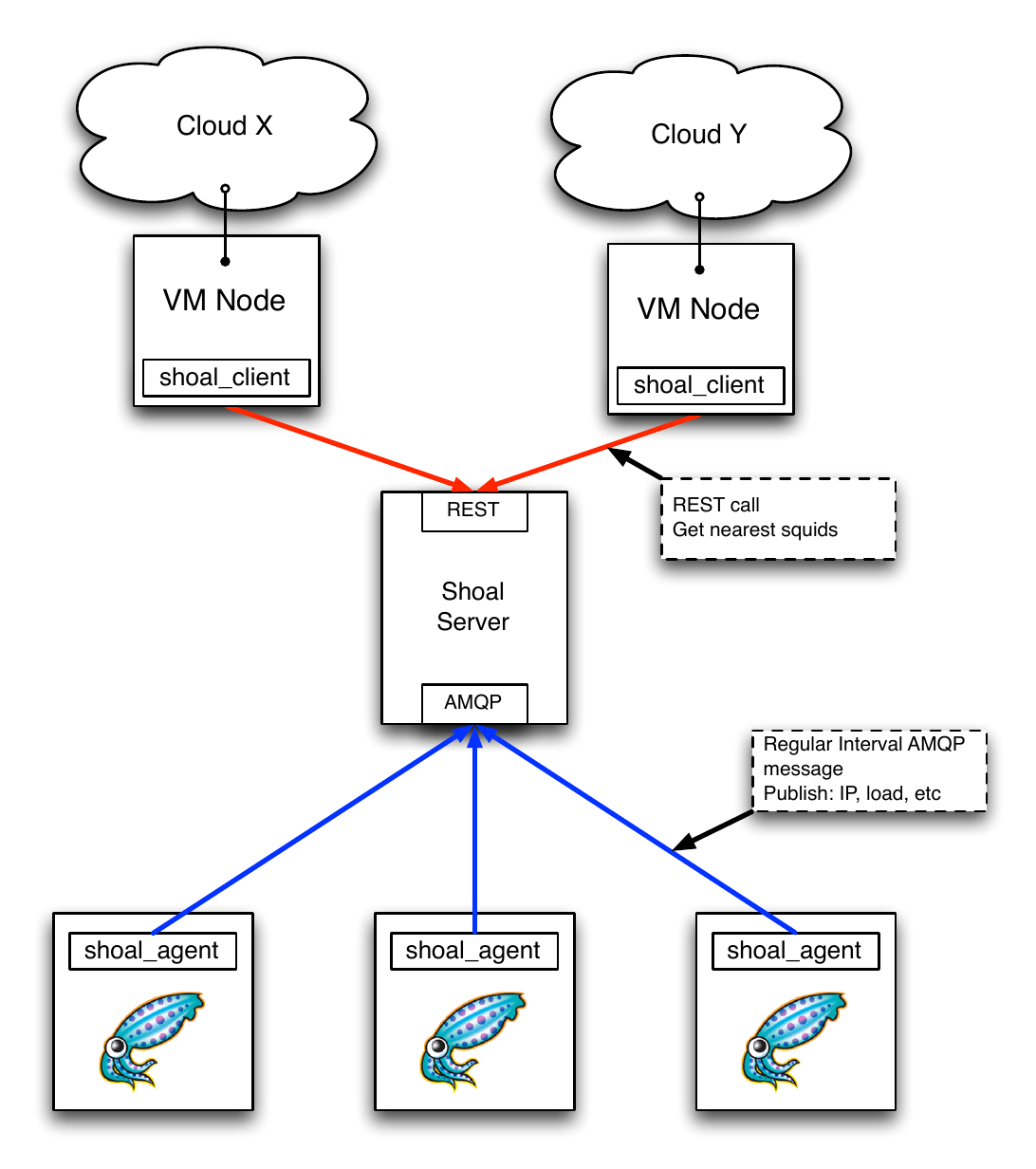}
\caption{Interaction between \shoal modules. \shoal Agents will send regular interval 
AMQP messages to \shoal Server. 
\shoal Clients residing on IaaS clouds will query the REST interface on \shoal Server to 
retrieve a list of the geographically nearest Squid servers.}
\label{shoal-image}
\end{center}
\end{figure}

Figure~\ref{shoal-image} shows an overview of \shoal and the interaction between each module. 
\shoal Server runs at a centralized location with a public IP address. 
For agents (i.e.\ Squid servers), \shoal Server will consume the heartbeat messages sent 
and maintain an up-to-date list of active Squids. 
For clients, \shoal Server will return a list of Squids organized by geographical distance and load. 
For regular users of \shoal Server, a web server is provided. 
The web server generates dynamic web pages that display an overview of \shoal. 
All of the tracked Squid servers are displayed and updated periodically on \shoal Server's 
web user interface, and all client requests are available in the access logs.

\begin{figure}
\begin{center}
\includegraphics[width=0.85\textwidth]{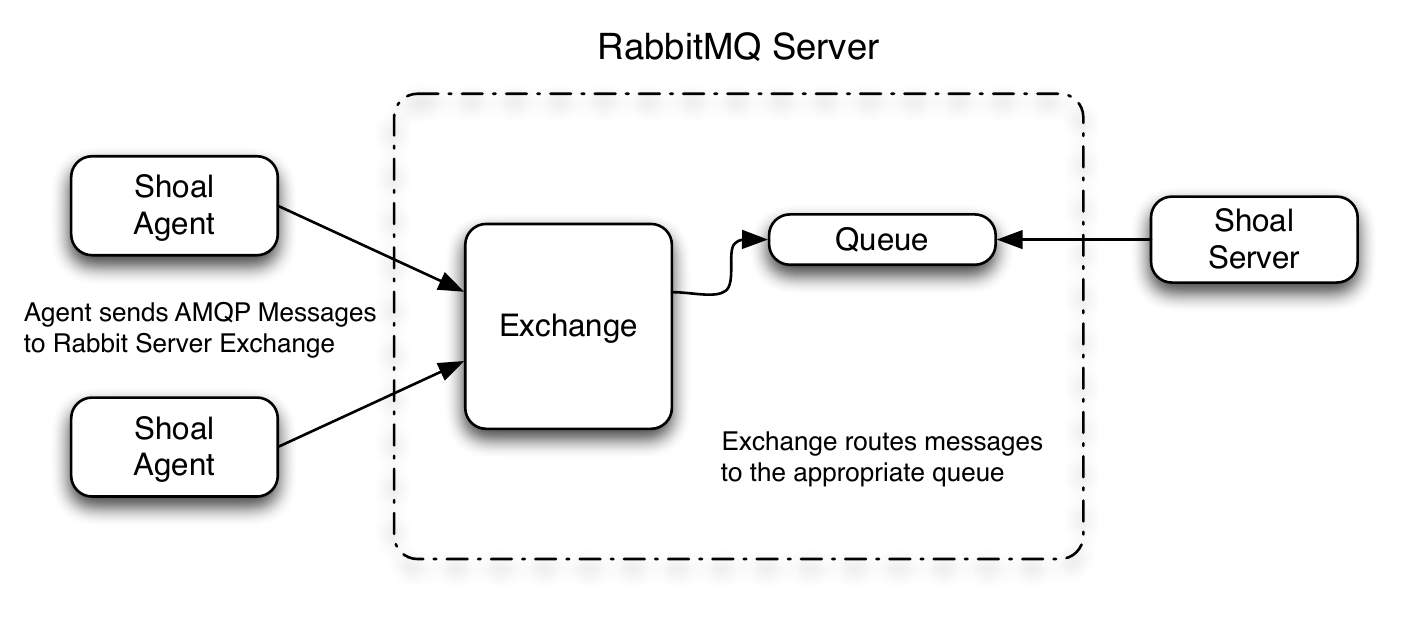}
\caption{RabbitMQ message routing between \shoal Agents and \shoal Server. 
All AMQP~\cite{ref:amqp} messages sent will be routed through the Exchange, and put 
into a queue. \shoal Server will then consume messages off this queue.}
\label{rabbitmq-flowchart}
\end{center}
\end{figure}

AMQP forms the communications backbone of \shoal Server. 
All information exchanges between \shoal Agent (Squid Servers) and \shoal Server are done 
using this protocol, and all messages are routed through a RabbitMQ~\cite{ref:rabbitmq} Server. 
Figure~\ref{rabbitmq-flowchart} gives a high level overview of message routing between 
\shoal Agents and \shoal Server.

The exchange will route the message according to a routing key that is set on the agent, 
and can be dynamic. 
The message will enter through the exchange and will then be delivered to a specific queue. 
\shoal Server creates this queue, and will receive all messages sent to the specific exchange. 
If \shoal Server is not running, all messages sent to the exchange will be discarded as there 
will be no queue for them to be routed to. 
More complicated routing can be done with the exchange-queue system, but currently it is a 
simple message hand-off from exchange to queue.


\section{Results}

\shoal was designed to handle large numbers of requests for the location of Squid caches
and to receive continuous updates from many Squid caches. 
AMQP was selected for the method of communication between \shoal Agent and \shoal Server 
because of its robustness and the possibility of using message queues to scale system 
components horizontally. 


\begin{figure}[t]
\begin{center}
\includegraphics[width=14cm]{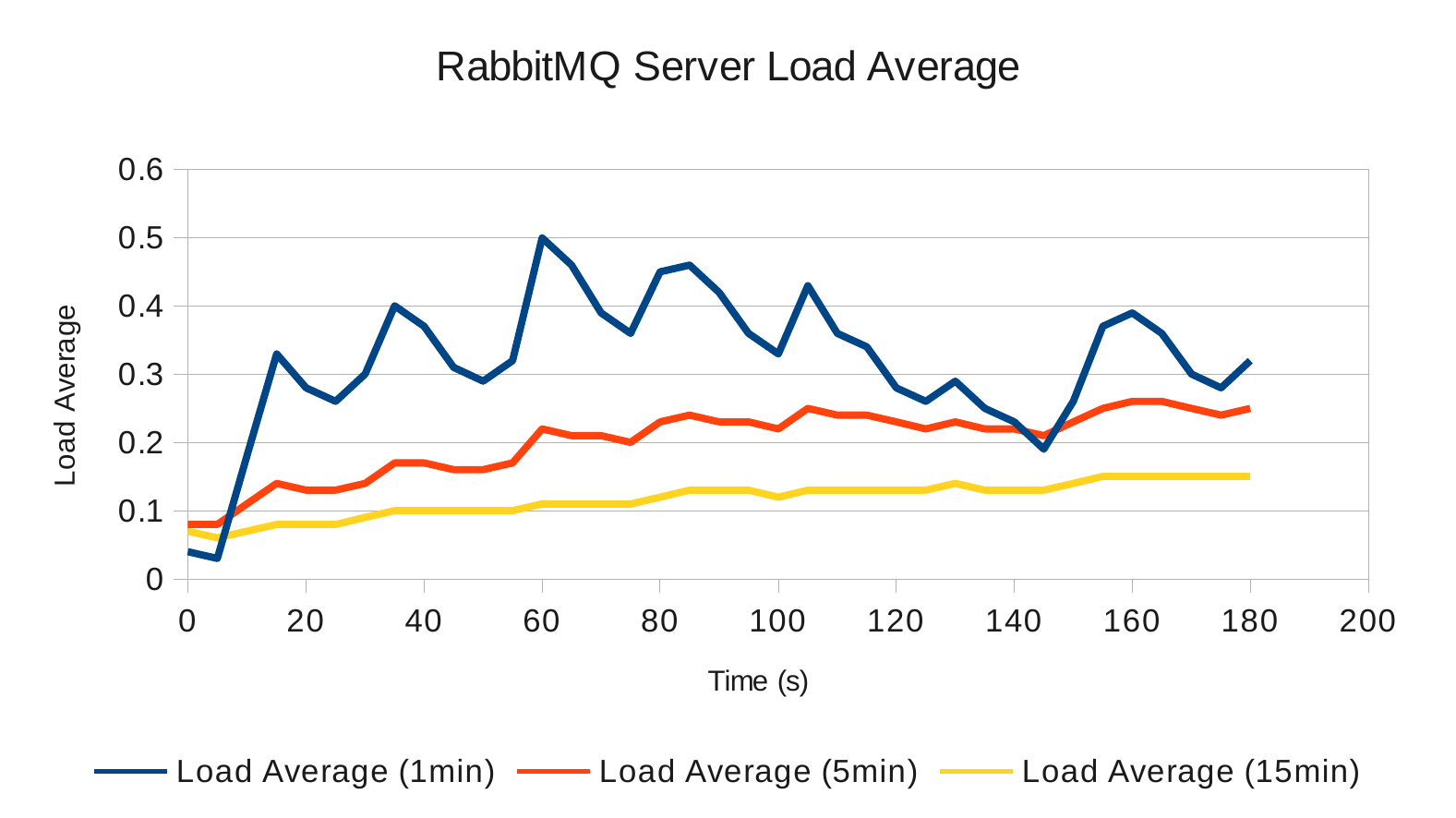}
\caption{RabbitMQ server load average when it is receiving 10000 AMQP messages/minute.
The load average is measured every 10 seconds over a 3 minute period.}
\label{shoalserver-loadavg-5000squid}
\end{center}
\end{figure}

In order to ensure that the AMQP messaging system would work effectively, a RabbitMQ server 
was configured on a VM running Scientific Linux~6.3~\cite{ref:scientificlinux} with 
1GB of RAM and a single processing core. 
The server was sent 10000 \shoal Agent messages per minute and the load 
average~\cite{ref:loadavg} was measured over a period of three minutes. 
This is equivalent to 5000 Squid Agents contacting the \shoal Server every 30 seconds. 
Figure \ref{shoalserver-loadavg-5000squid} shows that the system load average during 
the test was consistently below 0.5, 
showing that RabbitMQ can handle a significant load with minimal resources.

In order to ensure that \shoal is able to function at the scale required, a number of 
benchmark tests were performed. 
The \shoal Server was set up on a 16 core x86\_64 machine with 64 GB of RAM. 
A RabbitMQ server was set up on the previously described Virtual Machine. 
A client machine with the Apache benchmarking tool {\it ab}~\cite{ref:abtool} was set 
up to access the \shoal Server with a configurable number of parallel accesses, 
thus simulating the load from many worker nodes.


\begin{figure}[t]
\centering
\begin{minipage}{14cm}
\centering
\includegraphics[width=14cm]{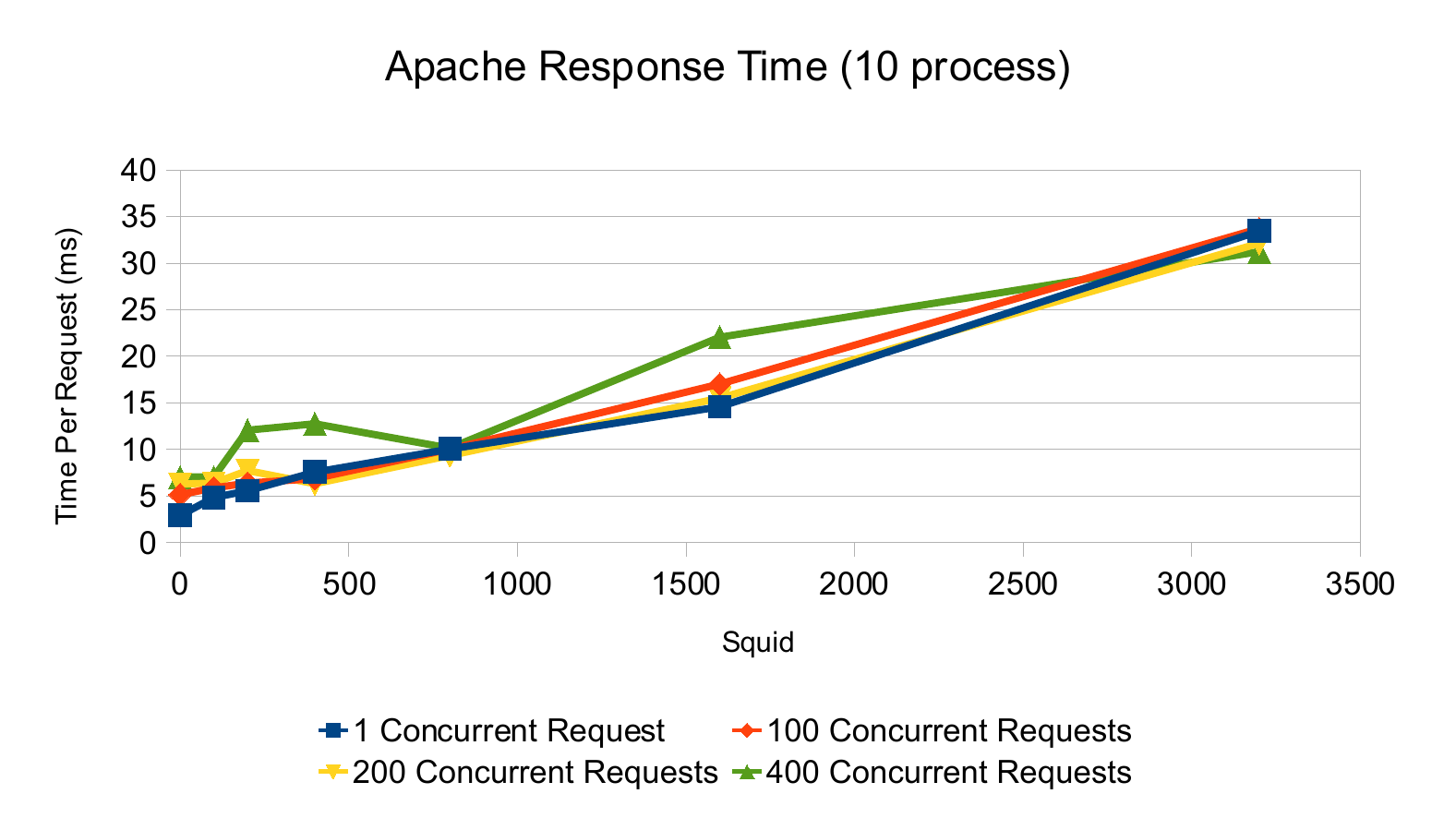}
\end{minipage}\hspace{2pc}%
\\
\centering
\begin{minipage}{14cm}
\centering
\includegraphics[width=14cm]{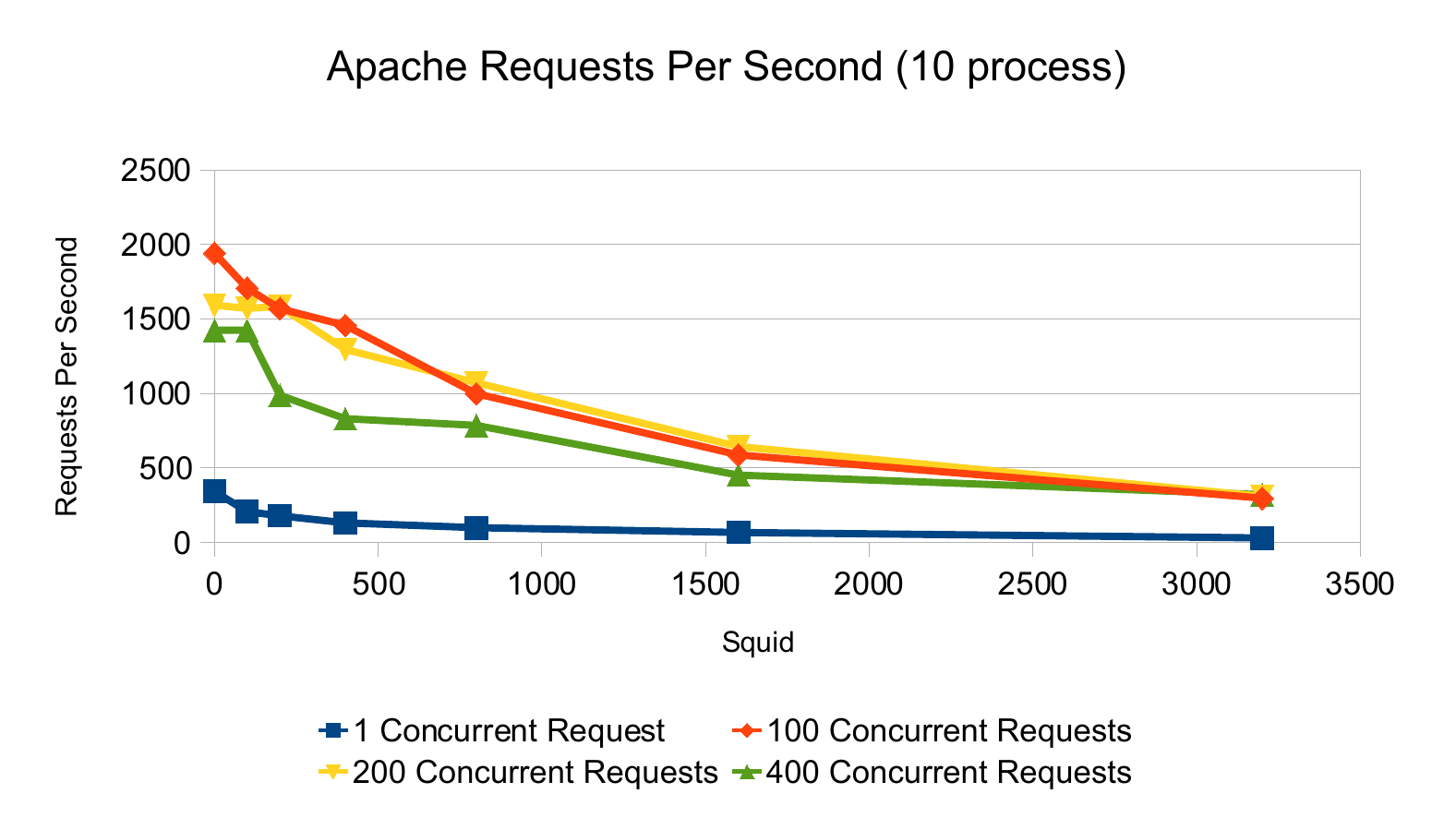}
\end{minipage}
\caption{The top plot shows the response time of the \shoal Server shown as a function 
of the number of  Squid caches for 1, 100, 200 and 400 concurrent requests.
The bottom plot shows the number of requests per second that the \shoal Server can sustain
as a function of the number of Squid caches for 1, 100, 200 and 400 concurrent requests.}
\label{apache-rt-10}
\end{figure}

The top plot in figure~\ref{apache-rt-10} shows the response time of the \shoal Server 
as a function of the number of Squid servers. 
The test was conducted using 1, 100, 200, and 400 concurrent connections. 
We observed that the response times increased roughly linearly with the number of Squid caches, 
with an approximate slope of 10 ms per 1000 Squid caches. 

The bottom plot of figure~\ref{apache-rt-10} shows the same benchmark with the data 
plotted as total number of requests which can be satisfied per second. 
We see that a single \shoal Server, tracking 800 squid servers, is able to respond 
to 1000 requests per second. 
This is equivalent to handling 1.8 million worker nodes requesting their nearest Squid 
Proxy every 30 minutes, demonstrating that \shoal has sufficient scalability for large-scale 
deployments.

\section{Future Work}
The algorithm for finding the nearest Squid server can be substantially improved if a 
temporary cache is used for different IP subnets. 
This would allow \shoal Server to use a hash table to store the nearest Squid servers, reducing 
the majority of RESTful API calls to complete in $O(1)$ time, whereas in its current 
implementation each web request requires $O(nlog(n))$ time, where \textit{n} is the 
number of tracked Squids. 

Work is underway to authenticate the Squid Caches registering with the \shoal Server 
using Secure Socket Layer and X.509~\cite{ref:rfc3820} certificates. 
This measure is necessary to prevent malicious Squids from advertising themselves to the 
\shoal server. 
However, it is important to note that the CVMFS repositories are cryptographically signed, 
so malicious Squid caches pose a denial of service risk for CVMFS rather than a 
code injection risk.

\pagebreak
\section{Conclusion}
We have developed a highly scalable application, called \shoal, for tracking and utilizing 
a distributed set of Squid HTTP web caches.
We see that a single \shoal Server tracking 800 squid servers is able to respond to 1000 
client requests per second. 
This is equivalent to handling over 1 million worker nodes requesting their nearest 
Squid Proxy every 30 minutes, demonstrating that \shoal can scale for massive deployments.

\section{ Acknowledgments }
The support of CANARIE and the Natural Sciences and Engineering Research Council of 
Canada are acknowledged.

\end{document}